\documentclass[preprint,12pt]{elsarticle}
\usepackage[T1]{fontenc}
\usepackage{textcomp}
\usepackage{paralist}
\usepackage{graphicx}      
\usepackage{natbib}        
\usepackage[cmex10]{amsmath}
\usepackage{amsfonts}
\usepackage{amssymb}
\usepackage{amsmath}
\usepackage{pgfplots}
\usepackage{url}
\pgfplotsset{compat=1.5}
\usepackage{array}
\usepackage{color,soul}
\usepackage{amsthm}
\usepackage{setspace}
\usepackage{hyphenat}

\newtheorem{Def}{Definition}
\DeclareMathOperator*{\argminB}{argmin}

\journal{Mechanical Systems and Signal Processing}

\begin{document}

\begin{frontmatter}

\author[Add]{Rishi Relan\corref{cor1}\fnref{fn1}} 
\ead{rishi.relan@vub.ac.be}
\cortext[cor1]{Corresponding author}
\fntext[fn1] {This work was supported in part by the IWT-SBO BATTLE 639,  Fund for Scientific Research (FWO-Vlaanderen), by the Flemish Government (Methusalem), the Belgian Government through the Inter university Poles of Attraction (IAP VII) Program, and by the ERC advanced grant SNLSID, under contract 320378.}

\author[Add]{Koen Tiels \fnref{fn1}}

\author[Add]{Anna Marconato \fnref{fn1}}

\author[Add]{Philippe Dreesen \fnref{fn1}} 

\author[Add]{Johan Schoukens \fnref{fn1}}

\address[Add]{Vrije Universiteit Brussel, Department ELEC \\ Pleinlaan 2, , 1050 Brussels, Belgium}                                              

\title{Data Driven Discrete-time Parsimonious Identification of a Nonlinear State-Space Model for a Weakly Nonlinear System with Short Data Record  \\ \linespread{1.5}
 \footnotesize{\textcolor{blue}{ArXiV Preprint: The original article is published in the journal of Mechanical Systems and Signal Processing (MSSP), Volume 104, 2018, Pages 929-943 \\ DOI: 10.1016/j.ymssp.2017.09.015}}}


\author{}

\address{}

\begin{abstract}
Many real world systems exhibit a quasi linear or weakly nonlinear behavior during normal operation, and a hard saturation effect for high peaks of the input signal.  In this paper, a methodology to identify a parsimonious discrete-time nonlinear state space model (NLSS) for the nonlinear dynamical system with relatively short data record is proposed. The capability of the NLSS model structure is demonstrated by introducing two different initialisation schemes, one of them using multivariate polynomials. In addition, a method using first-order information of the multivariate polynomials and tensor decomposition is employed to obtain the parsimonious decoupled representation of the set of multivariate real polynomials estimated during the identification of NLSS model. Finally, the experimental verification of the model structure is done on the cascaded water-benchmark identification problem. 

\end{abstract}

\end{frontmatter}


\section{\textbf{Introduction}}
There is an evident need of good system modelling techniques in many branches of engineering. Mathematical (linear or nonlinear) models are needed in various applications, for example, to understand and analyse the system under test, to simulate or predict the behavior of the system during the design phase or to design and implement a controller. System identification provides us with a variety of methods to derive accurate mathematical descriptions of the underlying system, based on a set of input/output measurements. Amount and quality of data plays an important role in any system identification framework. In some cases, increasing the measurement time is either not possible for example, when the input cannot be chosen and the system is unstable, the output can be exponentially growing. This essentially restricts the measurement time. Similarly, a major challenge is present in the study of systems whose behavior varies nonlinearly with time or task, resulting in small data records, or data that can be considered stationary for only short periods of time.

\subsection{\textbf{Nonlinear System Identification}}
The recent years have witnessed the shift from linear system identification \cite{ljung1998, RikJohanBook2012, van2012subspace}  to  nonlinear  system identification methods,  driven  by  the  need  to  capture  the inherent  nonlinear  effects  of  real-life  systems \cite{Eriten2013245, Feldman201765, Tiso2017185, Noel20172, Noel2017171}.  Nonlinear system identification constantly faces the challenge of deciding between the complexity of the fitted model and its parsimony. Flexibility refers to the ability of the model to capture complex nonlinearities, while parsimony is its ability to possess a low number of parameters. 

A  general framework  for  nonlinear  system  identification  does  not exist \cite{Giannakis2001533}, however, modeling nonlinear systems is covered in different fields like statistical learning and machine learning \cite{suykens2002least, rasmussen2006gaussian, hastie2009elements, suykens2012artificial, Cheng2017340}, but most of these methods are typically not specifically developed to deal with dynamics and often have limited means for dealing with noise. Within the system identification community two major approaches to nonlinear system identification can be distinguished: black-box nonlinear system identification \cite{Sjoberg19951691, young2011recursive, billings2013nonlinear} and block-oriented system identification \cite{giri2010block, mzyk2013combined}.       

State-space models are general representations that allow one to describe a variety of systems. In particular, nonlinear state-space modeling represents a promising, and at the same time challenging, class of techniques. In this paper, we focus mainly on black-box identification of nonlinear state space model (NLSS) structures \cite{Paduart2010, schon2011system}. 

The main contribution of this paper is the proposal of a data-driven nonlinear modelling methodology based on the initialisation methods proposed by \cite{Paduart2010, Anne2014} and the decoupling method proposed by \cite{Philippe2015Decouple}, for the identification of nonlinear state-space models for the cascaded water-tanks benchmark problem \cite{SchoukensM2016b}.  The effect of various factors affecting the suitability and the performance of these methods to capture the dynamical behaviour of the cascaded water-tanks benchmark problem is also discussed.

This paper is organised in the following Sections: Section \ref{CasWT} describes an example of weakly nonlinear system i.e. cascaded water-tanks benchmark, identification challenges associated with this benchmark problem. Section \ref{ResDis} gives an introduction to the experimental set-up. Thereafter the measurement methodology used for the acquisition of the signals is discussed. Section \ref{NonMod} describes the nonlinear modelling approach using the NLSS model structures used in this paper. 
The identification of NLSS model along with two different initialisation schemes is described Section \ref{IniSchemes}. Section \ref{FullNLSS} provides an overview of the final objective functions, which are minimised using two different initialisation schemes. The method to obtain the parsimonious representation of Polynomial Nonlinear State Space model is presented in Section \ref{sec:CPD}. Results are presented in Section \ref{Results}, and finally, the conclusions  are given in Section \ref{Conc}.     

\section{\textbf{Cascaded Water Tanks System}}
\label{CasWT}

\begin{figure}[!ht]
\centering
\includegraphics[width=0.375\textwidth]{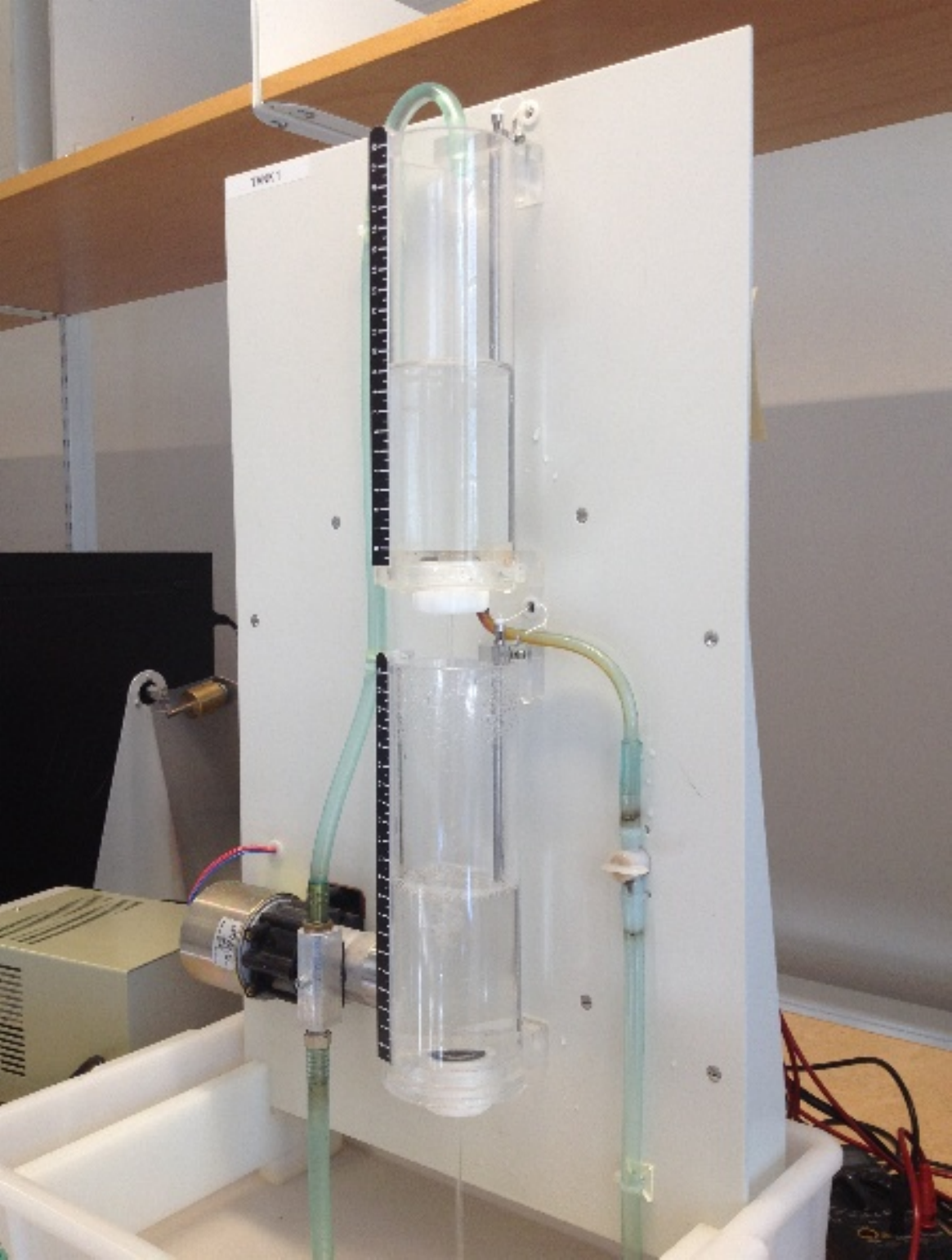}
\caption{ The water is pumped from a reservoir in the upper tank, flows to the lower tank and  finally  flows back into the reservoir.  The input is the pump voltage, the output is the water level of the lower tank.}
\label{WaterTanks}
\end{figure}

The cascaded tanks system is a  liquid level control system consisting of two tanks with free outlets fed by a pump.  The input signal controls a water pump that pumps the water
from  a  reservoir  into  the  upper  water  tank.   The  water  of  the  upper  water  tank   flows
through a small opening into the lower water tank, and  finally through a small opening
from the lower water tank back into the reservoir. The relation between \begin{inparaenum} \item  the water  flowing from the upper tank to the lower tank and \item the water  flowing from the lower tank into the reservoir are weakly nonlinear functions. \end{inparaenum} 

However, when the amplitude of the input signal is too large, an overflow can happen in the upper tank, and with a delay also in the lower tank.  When the upper tank overflows,
a part of the water goes into the lower tank, the rest  flows directly into the reservoir.  This effect is partly stochastic, hence it acts as an input-dependent process noise source.  Without  considering  the  overflow  effect,  the  following  input-output  model  can  be  constructed based on Bernoulli's principle and conservation of mass \cite{SchoukensM2016b}:
\begin{align}
 \dot{x_1}(t) &= -k_1 \sqrt{x_1(t)} + k_4 u(t) + w_1(t), \nonumber \\ 
 \dot{x_2}(t) &=  k_2 \sqrt{x_1(t)} -k_3 \sqrt{x_2(t)} + w_2(t),\nonumber \\
 y(t) &= x_2(t)+ e(t)
\label{eqn:WT}
\end{align} where $u(t)$ is the input signal, $x_1(t)$ and $x_2(t)$ are the states of the system, $w_1(t)$, $w_2(t)$ and $e(t)$ are the additive noise sources and $k_1, k_2, k_3,$ and $k_4$ are the constants depending on the system properties. 

\begin{figure}[!ht]
\centering
\includegraphics[width=1\textwidth]{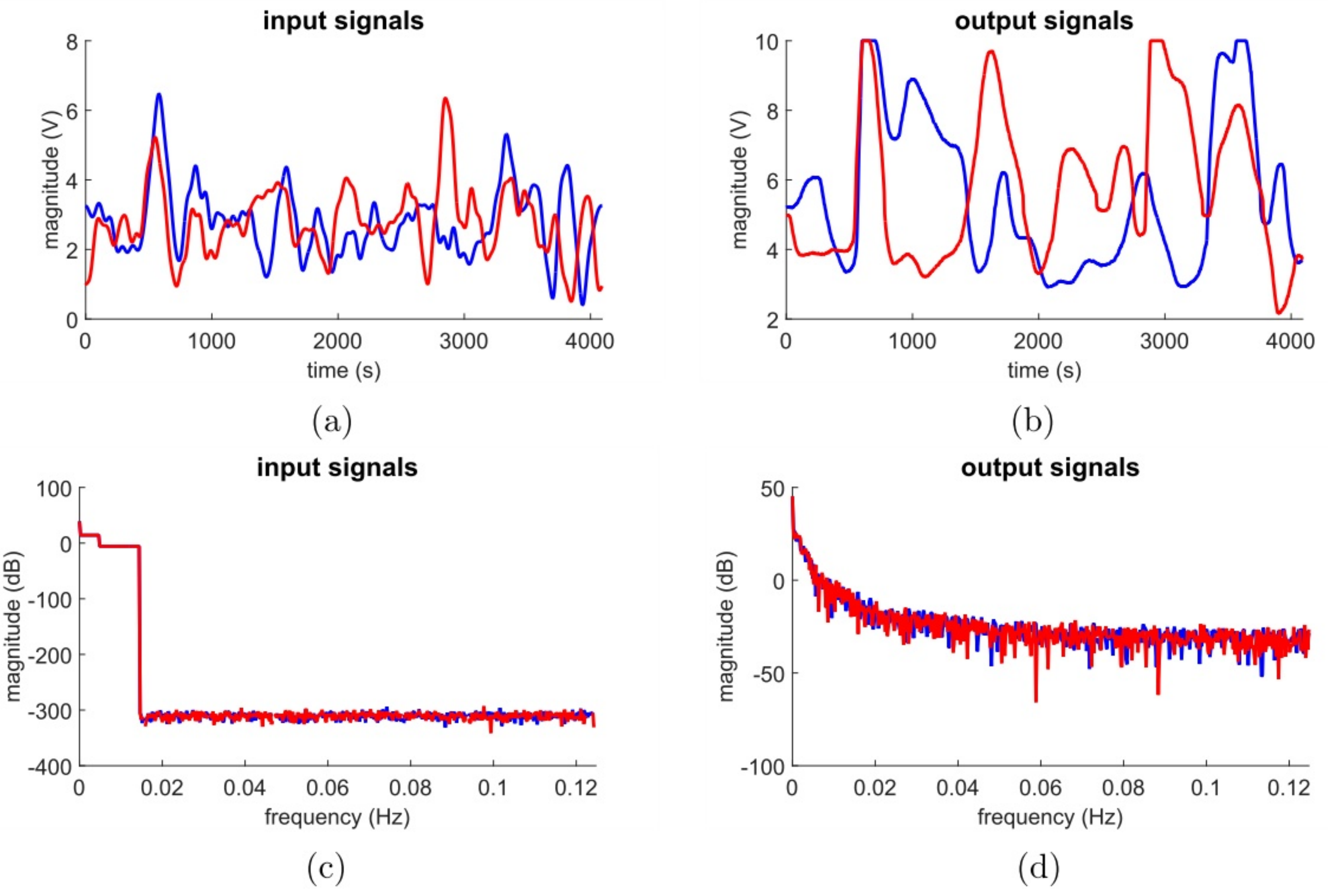}   
\caption{Input (a,c) and output (b,d) signals of the estimation (blue) and test (red) data
records in the time (a,b) and frequency (c,d) domain.}
\label{InOutWT}
\end{figure} 
 
\subsection{Measurement Set-up}
\label{ResDis}
In this section, we describe the measurement set-up and data acquisition procedure which was employed to acquire the input-output signals from the benchmark system. The input signals are multisine signals which are defined as:
 \begin{equation}
u_{ms}(t)= \sum\limits^{k_{max}}_{k=1}{A}(k) \cos(k \hspace{0.09cm} 2 \hspace{0.09cm} \pi \frac{f_s}{N} \hspace{0.09cm} t+\varphi_{k})
\label{eqn:mls1}
\end{equation} where $f_{max}= k_{max}\frac{f_s}{N}= 0.0144$ Hz. The period length of the  excitation signal is $1024$ points, the lowest frequencies have a higher amplitude then the higher frequencies (\cite{SchoukensM2016b}, see figure \ref{InOutWT}).  The sample period $T_s$ is equal to $4$s.  The input signals are zero-order hold input signals.

The process is controlled from a Matlab interface to the A/D and D/A converters attached to  the  water  level  sensor  and  the  pump  actuator.   The  water  level  is  measured  using capacitive water level sensors,  the measured output signals have a signal-to-noise ratio that is close to $40$ dB. The water level sensors are considered to be part of the system, they are not calibrated and can introduce an extra source of nonlinear behavior.  The system states have an unknown initial value at the start of the measurements.  This unknown state is the same for both the estimation and the test data record.

\subsection{Identification Challenges}
The major  nonlinear system identification challenges associated with the water-tanks benchmark are:
\begin{inparaenum}
\item the hard saturation nonlinearity combined with the weakly nonlinear behavior of the system in normal operation,
\item the overflow from the upper to the lower tank, this effect also introduces input dependent process noise,
\item the relatively short estimation data record,
\item the unknown initial values of the states.
\end{inparaenum} 

\section{\textbf{Generic Nonlinear State Space Model Structure}}
\label{NonMod}
A physical interpretation of the system under test is not always required, for instance in control or prediction problems. In that case, the user prefers an easy-to-initialise black-box model. Moreover, the model should preferably be able to describe Multiple-Input Multiple-Output (MIMO) systems in a compact way. A good base for such a model is a state space representation of  the  system under consideration.  A  general ${n^{th}}$ order  discrete-time state space   model   is   described   by   the following equations:
\begin{align}
 x(t+1) &= f(x(t),u(t))\nonumber \\ 
 y(t) &= g(x(t),u(t))
\label{eqn:NSS}
\end{align}with $u(t)\in \mathbb{R}^{n_u}$  the vector containing the ${n_u}$ inputs at time $t$, and $y(t)\in \mathbb{R}^{n_y}$  the vector containing the ${n_y}$ outputs. The state vector $x(t)\in \mathbb{R}^{n}$  represents the memory of the dynamical system.

\subsection{\textbf{Polynomial Nonlinear State Space Model}}
\label{PLNSS}
A nonlinear state space model (where $f(\cdot), g(\cdot)$ are approximated by polynomial basis functions) is termed as a Polynomial Nonlinear State-Space (PNLSS) model. The PNLSS model structure \cite{Paduart2010} is described as:
\begin{align}
 x(t+1) &= Ax(t) + Bu(t) + E\zeta(t)\nonumber \\
 y(t) &= Cx(t) + Du(t) + F\eta(t)+e(t)
\label{eqn:PLNSS}
\end{align}The coefficients of the linear terms in $x(t)\in \mathbb{R}^{n}$   and $u(t)\in \mathbb{R}^{n_u}$  are given by the matrices $A \in \mathbb{R}^{n \times n}$ and $B \in \mathbb{R}^{n \times n_u}$ in the state equation,  $C \in \mathbb{R}^{n_y \times n}$ and $D \in \mathbb{R}^{n_y \times n_u}$ in the output equation. The  vectors $\zeta(t) \in \mathbb{R}^{n_{\zeta}}$ and $\eta(t) \in \mathbb{R}^{n_{\eta}}$ contain nonlinear monomials in $x(t)$ and $u(t)$ of  degree  two  up  to  a  chosen  degree  $d$ and $e(t)$ is the measurement noise.  The coefficients associated with these nonlinear terms are given by the matrices $E \in \mathbb{R}^{n \times n_{\zeta}}$ and $F \in \mathbb{R}^{n_y \times n_{\eta}}$. 

\subsection{\textbf{Nonlinear State Space Model-2}}
\label{NLSS2}
Another way (hereinafter termed as NLSS2) of describing the nonlinear state space model structure in (\ref{eqn:NSS}) is by describing it in the form as described in \cite{marconato2014improved}:
\begin{align}
 x(t+1) &= Ax(t) + Bu(t) + f_{NL}(x(t), u(t))\nonumber \\
y(t) &= Cx(t) + Du(t) + g_{NL}(x(t), u(t))+e(t)
\label{eqn:NLSSIMP}
\end{align} where $A \in \mathbb{R}^{n \times n}$ and $B \in \mathbb{R}^{n \times n_u}$ in the state equation,  $C \in \mathbb{R}^{n_y \times n}$ and $D \in \mathbb{R}^{n_y \times n_u}$ in the output equation. The nonlinear functions $f_{NL}(\bullet)$ and $ g_{NL}(\bullet)$ have $n$ and $n_y$ outputs respectively. The next section discusses the steps involved in the identification of PNLSS and NLSS2 models from input-output measurements

\subsection{\textbf{Identification of the NLSS}}
\label{IdenPNLSS}
Here, we describe first the work-flow associated with the identification of NLSS model structure. 
\begin{figure}[!ht]
\includegraphics[width=\textwidth]{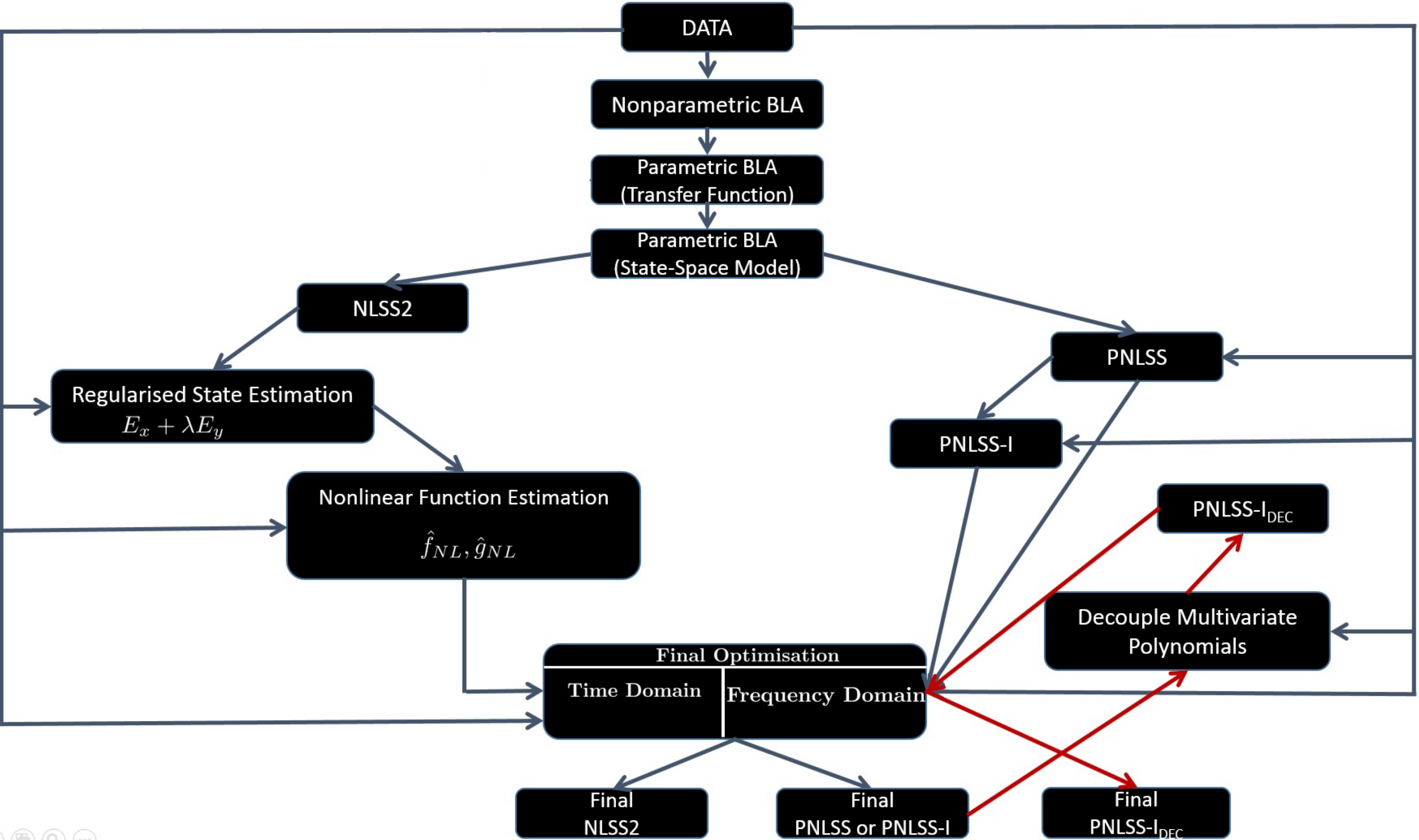}   
\caption{Workflow of the identification procedure for different NLSS model structure}
\label{fig:WorkFlow}
\end{figure} 
The overall identification workflow for different NLSS model structures described in this paper consists of the steps shown in figure \ref{fig:WorkFlow}. The structure of the black-box NLSS model given in (\ref{eqn:PLNSS}) or (\ref{eqn:NLSSIMP}) lends itself to an efficient, four major steps identification procedure. 
\begin{enumerate}
 \item First, initial estimates of the $A, B, C$ and $D$ matrices are obtained. In order to do so, first, a nonparametric estimate of the system's frequency response function (FRF) is determined in mean square sense. Then, a parametric linear model (linear subspace $A,B, C, D$ matrices) is estimated  from this nonparametric Best  Linear  Approximation (BLA).
 \item  Thereafter depending on the model structure used, the model is initialised either with an estimate of only the linear subspace $A,B, C, D$ matrices (PNLSS model structure) or $A,B, C, D$ along with an estimate of nonlinear functions $f_{NL}$ and $g_{NL}$ (NLSS2 model structure).
\item Finally, the last step consists in estimating the full nonlinear model by using again a nonlinear search routine namely the Levenberg-Marquardt method \cite{levenberg1944, Nocedal2006NO}. 
 \item  Once the PNLSS model is obtained, then a decoupled representation of the multivariate polynomials in state and output equations is  obtained (in a static dimensionality reduction step), before plugging back the decoupled representation into the model structure and doing the final optimisation (see the red arrows in Fig. \ref{fig:WorkFlow}).
\end{enumerate}

\begin{figure}[!ht]
\centering
\includegraphics[scale=0.67, trim = {1.75cm 5cm 2cm 2.5cm},clip]{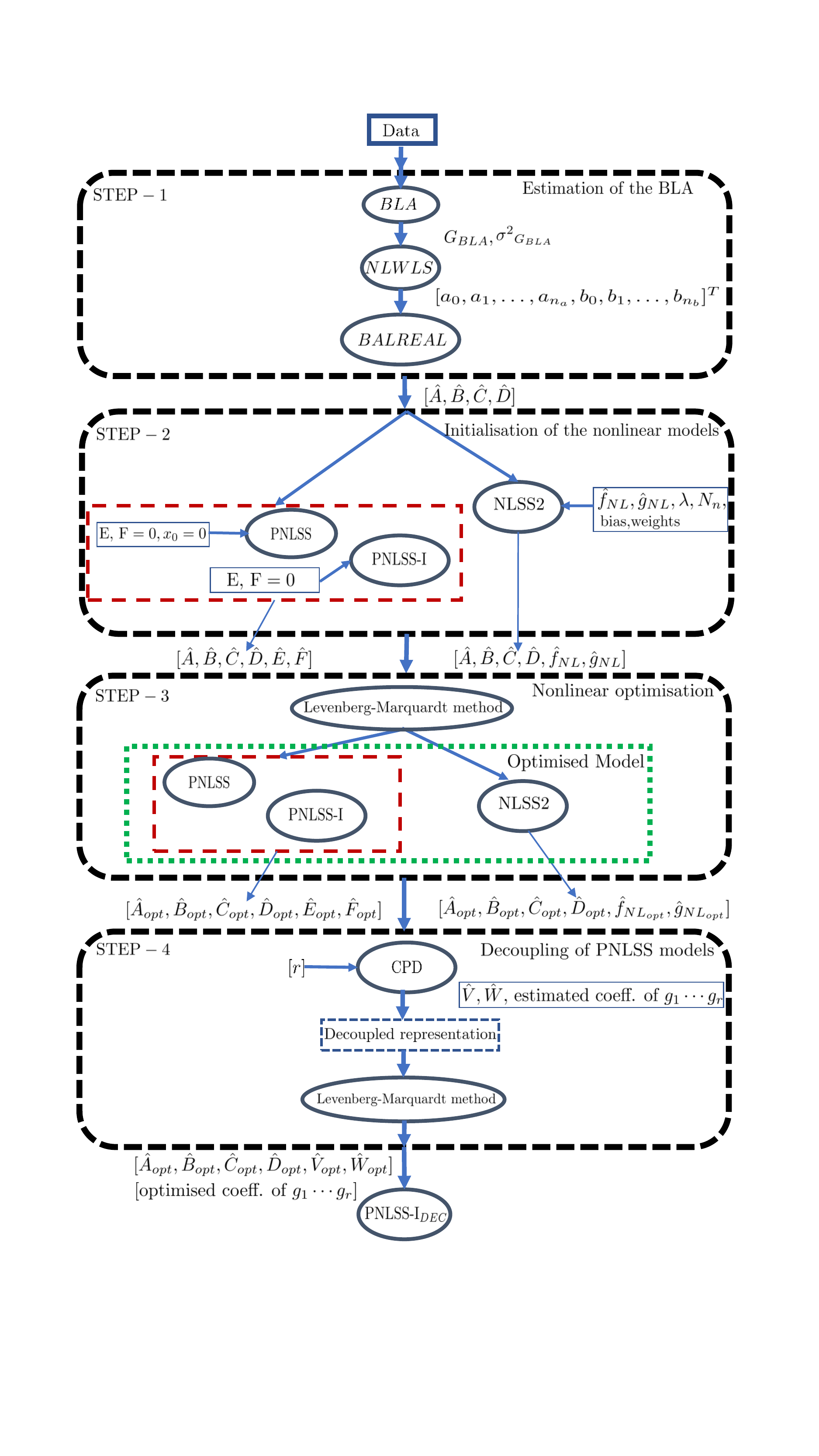}   
\caption{Algorithmic work flow of the identification procedure.}
\label{fig:WorkFlowAlgo}
\end{figure} Algorithmic work flow for the four steps described above is shown in Figure. \ref{fig:WorkFlowAlgo}.It should be noted that bounded input-bounded output (BIBO) stability is required for this optimization procedure. Note that the proposed approach targets systems for which the dynamics  can  be  captured  by  the  BLA  and  systems  that  are assumed to have only one equilibrium point. In the subsections below, these steps as well as the framework involved in these steps are described briefly. 
\subsubsection{\textbf{Best Linear Approximation}}

\begin{Def}   
	The Best Linear Approximation (BLA) of a nonlinear system is defined as the model $G$ belonging to the set of linear models $\mathcal{G}$, such that
\begin{equation}
 G_{BLA}= \underset{G \in \mathcal{G}}{\operatorname{arg\,min}} \hspace{0.1cm} \mathbb{E} \left( |y(t)-Gu(t)|^2 \right)
\end{equation}
\end{Def} 

Within the above described set up, the nonparametric BLA and the variance of the BLA (${{\sigma}^2}_{G_{BLA}}$) can be calculated very efficiently in mean square sense, using either the Fast or the Robust method explained in \cite{RikJohanBook2012}, or the Local Polynomial Method (LPM) \cite{RikJohanBook2012, Schoukens2009260}.
\subsubsection{\textbf{Parametric BLA}}
A parametric model is more convenient for simulation and control purposes. It offers extra opportunities to better understand the system behaviour using the pole-zero representation. Hence, our next goal is to obtain a parametric  model of the system under consideration. Thus, using the nonparametric FRF estimate ($G_{BLA}$) and its variance (${{\sigma}^2}_{G_{BLA}}$), which is found in the previous step using the LPM approach \cite{RikJohanBook2012, Schoukens2009260},  we make a parametric model of our system by solving a weighted least squares problem. This model (discrete-time) describes the system as a rational transfer function. The model considered here is a rational function in the backward shift operator $q^{-1}$:
\begin{equation}
 \hat{G}_{BLA}(q,\theta_{tf}) = \frac{b_0+b_1q^{-1}+b_2q^{-2}+......+b_{n_b}q^{-n_b}}{a_0+a_1q^{-1}+a_2q^{-2}+......+a_{n_a}q^{-n_a}},
\label{eqn:paraEq}
\end{equation} The parameter vector $\theta_{tf} \in \mathbb{R}^{(n_b+n_a+2) \times 1}$  contains the parameters  $[a_{0},a_{1},\ldots , \\ a_{n_a}, b_{0},b_{1},\ldots ,b_{n_b}]^T$. Since one parameter can be chosen freely because of the scaling invariance of the transfer function, only $n_b + n_a + 1$ independent parameters need to be estimated by minimizing a nonlinear weighted least squares (NLWLS) cost function:

\begin{equation}
V_{tf}(\theta_{tf}) = \sum\limits_{k = 1}^{F} \frac{|G_{BLA}(e^{j\omega_{k}} ) - \hat{G}_{BLA}(e^{j\omega_{k}},\theta_{tf})|^2}{{\sigma_{BLA}^2}(e^{j\omega_{k}})},
\end{equation} 
where $G_{BLA}(e^{j\omega_{k}} )$ is the frequency domain representation of \eqref{eqn:paraEq} and ${\sigma_{BLA}^2}(e^{j\omega_{k}})$ includes both noise and nonlinear distortion. The order of the parametric model in (\ref{eqn:paraEq}) can for example be determined using a signal theoretic measure such as the minimum description length (MDL) criterion (see pp. no. $439$ of \cite{RikJohanBook2012}). 
This NLWLS framework also guarantees the lowest possible uncertainty on the model parameters, i.e. the efficiency of the estimates \cite{RikJohanBook2012}.  

By using the function \textit{Balreal} from the \textit{Control System Toolbox} of MATLAB \cite{MATLAB2013b}, a balanced state space realization $G_{ss} = (A,B,C and D)$ for the stable portion of the linear system $\hat{G}_{BLA}(q,\theta_{tf})$ is calculated. For stable systems, this is an equivalent realization for which the controllability and observability Gramians are equal and diagonal \cite{moore1981principal,Laub1987}. The resulting estimate of the state space matrices are $\hat{A}_{BLA}, \hat{B}_{BLA}, \hat{C}_{BLA}$ and $\hat{D}_{BLA}$, which is then used further in the PNLSS model structure for the initialisation of the linear part.

\section{\textbf{Initialisation of the NLSS model}}
\label{IniSchemes}
As explained earlier, the main difference between the two approaches discussed in Section \ref{NonMod} lies in the way, in which the full model structure in equations (\ref{eqn:NSS}) is initialised. In this section, we give an overview of the two different approaches followed in this paper. 

\subsection{Initialisation approach I}
In the first approach, the linear part of the PNLSS model structure (\ref{eqn:PLNSS}) is initialised using the BLA i.e the $\hat{A}_{BLA}, \hat{B}_{BLA}, \hat{C}_{BLA}, \hat{D}_{BLA}$ matrices obtained in the previous step. The coefficients of matrices $E$ and $F$ are initialised as numerical value $0$, but it is possible to select which of these coefficients are free during the final optimisation step by selecting e.g. either all, none or no cross-terms.  
\begin{align}
x(t+1) &= {A}\hspace{0.07em}x(t) + {B}\hspace{0.07em} u(t) + E\zeta(t)\nonumber \\
y(t) &= {C}\hspace{0.05em} x(t) + {D}\hspace{0.06em} u(t) + F\eta(t)+e(t)
\label{eqn:PLNSSExt} 
\end{align}

\subsubsection*{\underline{PNLSS-I Model (Estimation of Initial State $x_{0}$)}}
For both periodic and non-periodic excitations, there are two ways to estimate explicitly the initial state $x_0$ either to include it as an extra parameter in the final Levenberg-Marquardt optimization step of the full PNLSS model (step $3$ in Section \ref{IdenPNLSS}) or to estimate the initial conditions $x_{0}$, an extra column in the state space matrix $B$ and an extra value $1$ in the input vector is included (as the ordinary model parameters) \cite{Anne2014}. 

In this case study, we include  $x_{0}$ as an extra parameter to be optimised along with other parameters. Hereinafter whenever the initial condition is explicitly estimated, then the model is termed as PNLSS-I. For both PNLSS and PNLSS-I, we choose a $3^{rd}$ order model and the polynomial degrees up-to $3$.

\subsection{Initialisation approach II}
If the state sequence $x(t)$ would be exactly known, the problem of obtaining a nonlinear model could be solved much more easily by estimating   $f$ and $g$ individually and as static mappings \cite{marconato2014improved}. As the state sequence is not available in practice, one would like to obtain an approximation of $x(t)$ to be able to obtain initial estimates of $f_{NL}$ and $g_{NL}$. Hence the first step in this approach is to obtain an initial estimate of the state sequence $x(t)$.
\subsubsection{Estimation of $x_{LS}$}
Based on the obtained linear model $G_{ss}$, and on the set of available data $\{{u(t), y(t)}_{t=1}^{N}\}$ , an approximation of the (unknown) states $x(t)$ is obtained. If this can be done reliably, it is possible to solve an approximate version of (\ref{eqn:NSS}), eliminating the recursion in the state equation. An estimate of the states $\hat{x}_{LS}(t)$ is obtained by solving the following regularised least squares problem \cite{marconato2014improved} for $ t = 1,2.....,N$ :
\begin{align}
\label{eq:NLSS}
 \hat{x}_{LS}(t) &= \argminB_{x(t)} \sum\limits_{t} (y(t)- \hat{C}_{BLA} x(t))^2 \\ \nonumber
                 & + \lambda \sum\limits_{t} (x(t+1)-  \hat{A}_{BLA} x(t) - \hat{B}_{BLA} u(t))^2  \\ \nonumber
                 &= \argminB_{x(t)} E_{y}+ \lambda E_{x}
\end{align} The first term $E_{y}$ of the cost function represents the data fit, while the second term  $E_{x}$ represents the linear model fit; $\lambda$ is the tradeoff parameter that needs to be tuned to change the emphasis given on the two criteria. By tuning $\lambda$, a deviation from the linear state (resulting from the BLA estimates) is allowed. In practice, the optimal value of $\lambda$ is chosen such that  a  given  performance  criterion  is  optimized.  

The choice of $\lambda$ depends on the application and the final cost function.  In practice, $\lambda$ can  be learned by cross-validation. In  this  work, several initialised models resulting from different choices of $\lambda$ are compared, and the value of $\lambda$ minimising the RMSE between the initialised  model output and the measured output (for the validation data set) is selected. For an overview of the various factors (e.g. trade-off parameter $\lambda$ and number of neurons) affecting the performance of this model structure, the readers are referred to \cite{marconato2013study}. Alternatively, a \textit{Kalman Filter} formulation can also be used for the initialisation that would change the approximation slightly.

\subsubsection{Estimation of functions $f_{NL}$ and $g_{NL}$}
Once the estimate  ${\hat{x}_{LS}(t)}_{t=1}^{N}$ of the unknown state sequence is available, one obtains the following approximate static problem:
\begin{align}
\hat{x}_{LS}(t+1) &= f(\hat{x}_{LS}(t),u(t))+r_{LS}(t) \nonumber\\ 
                  &=  \hat{A}_{BLA}\hat{x}_{LS}(t)+ \hat{B}_{BLA}u(t) \nonumber \\ 
                    &  + f_{NL}(\hat{x}_{LS}(t),u(t))+r_{LS}(t) \label{eqn:XLS1} \\
y(t)              &=  g(\hat{x}_{LS}(t),u(t)) + e_{LS}(t) \nonumber \\ 
                  &=  \hat{C}_{BLA}\hat{x}_{LS}(t)+ \hat{D}_{BLA}u(t) \nonumber \\ 
                    &  + g_{NL}(\hat{x}_{LS}(t),u(t))+ e_{LS}(t) \label{eqn:XLS2}        
\end{align} where $r_{LS}(t)$ and $ e_{LS}(t) $ are error terms resulting from the fact that here the approximated state sequence is introduced in the problem. Equations \eqref{eqn:XLS1} and \eqref{eqn:XLS2} represent two static regression problems that can be solved independently employing simple regression methods. Note that at this stage, the recursion in the state equation is not present anymore, since the state sequence is now assumed to be known. Therefore, both functions $f_{NL}$ and $g_{NL}$ can be estimated as basis function expansions. 

The choice of basis functions is user defined and application dependent. User is free to select any smooth basis functions like neural networks, support vector machines, splines etc. for the estimation. Here we use the neural-networks with $N_n = 2$ hidden layer neurons and \textit{sigmoid} nonlinear function. These choices are inspired by the fact that the knowledge of model structure as well as nonlinearities can be extracted from (\ref{eqn:WT}). 

\section{Identification of full NLSS}
\label{FullNLSS}

In the last step, the coefficients of both the linear and the nonlinear terms in (\ref{eqn:PLNSS}) are identified. This problem can either be solved in frequency domain or time domain. In order to keep the estimates of the model parameters unbiased it is assumed that the input $u(t)$ of the model in Section \ref{NonMod} is noiseless, i.e., it is observed without any errors and it is independent of the output noise. It is worth noting that this assumption is violated in the cascaded water-tanks benchmark problem.

\subsection{\underline{PNLSS model}} A weighted least squares approach (in frequency domain) will be employed. The Weighted Least Squares (WLS) cost function that needs to be minimized with respect to the parameter $\theta_{NL} = [vec^{T}(A), vec^{T}(B), vec^{T}(C),$ $vec^{T}(D), vec^{T}(E),$ $vec^{T}(F)]^T$  is given by:
\begin{equation}
 V_{WLS}(\theta_{NL}) = \sum\limits_{k = 1}^{N_t} \frac{|Y_{mod}(j\omega_{k},\theta_{NL}) - Y(j\omega_{k})|^2}{W(j\omega_k)}
 \label{eqn:NonEst}
\end{equation} where $N_t$ is the total number of selected frequencies. $Y_{mod}$ and $Y$ are the DFTs of the modelled output and the measured output, respectively. Because in nonlinear systems, model errors often dominate the disturbing noise, we put the weighting factor $W(j\omega_k) = 1$. Only if the model errors are below the noise level, $W(j\omega_k) $ can be put equal to the noise variance ${\sigma_{n}}^2(j\omega_{k})$. Furthermore, model error $\epsilon (j\omega_{k},\theta_{NL}) \in \mathbb{C}^{n_y}$  is defined as
\begin{equation}
\epsilon (j\omega_{k},\theta_{NL}) = Y_{mod}(j\omega_{k},\theta_{NL}) - Y(j\omega_{k}).
\label{eqn:Error}
\end{equation} 

\subsection{\underline{NLSS2 model}} For the final optimisation of the model structure in (\ref{eqn:NLSSIMP}), the following cost function is solved, where the $\theta_{NL}$ now also contains the parameters of the basis functions $f_{NL}$ and $g_{NL}$.

\begin{equation}
 V_{LS}(\theta_{NL_2}) = {\frac{1}{N}}\sum_{t=1}^{N}(y(t)- \hat{y}(t,\theta_{NL_2}))^{2}
 \label{eqn:NonLSSEst}
\end{equation}

\section{Finding a Parsimonious Representation}
\label{sec:CPD}

Although PNLSS models (see Section \ref{FullNLSS}) are very capable to capture many nonlinear effects, one of the main disadvantages of the PNLSS model structure is that the number of parameters grow combinatorially  with  the  polynomial  degree  and the number of variables (both in inputs and states). It can be verified \cite{Paduart2010} that the total number of coefficients of nonlinear terms is given by 
\begin{equation} 
\left(
\frac{(n+n_u+d)!}{d\hspace{0.15em}! \hspace{0.375em}(n+n_u)!}- (n + n_u) \right)
\times (n+n_y),
\label{eq:Comb}
\end{equation}
where it is assumed that $\zeta$ and $\eta$ contain the same monomials, and both constants and linear terms are discarded. This is very disadvantageous especially in the case of short data records as it can lead to over-fitting or identifiability issues. Hence, in order to find a parsimonious representation of the full PNLSS model which contain multivariate polynomials $E \eta (t)$ and $F \zeta (t)$, in this section,  we use the  method proposed by \cite{Philippe2015Decouple} to decouple multivariate polynomials  in  the  PNLSS  model  into univariate polynomials. 

The decoupling method of \cite{Philippe2015Decouple} computes a decoupled representation (see Fig. \ref{fig:CPD}) of the coupled multivariate polynomials by means of the canonical polyadic decomposition (CPD) of a 3-dimensional matrix or tensor (see \cite{carroll1970analysis,harshman1970foundations, kolda2009tensor}). The advantage of this decoupling lies in the fact that it reduces the number of parameters to be estimated by removing the cross-terms in the multivariate polynomials. Later, the equivalent univariate parisimonious representation of the multivariate polynomials is plugged back into the PNLSS model structure to be further optimised w.r.t the output error.  
\subsection{A decoupled PNLSS model}
The polynomial decoupling method results in a new polynomial nonlinear state-space as 
\begin{eqnarray}
x(t+1) &=& Ax(t) + B u(t) + W_x g \begin{pmatrix}V^T \begin{bmatrix} x(t) \\ u(t) \end{bmatrix} \end{pmatrix}, \nonumber \\
y(t) &=&   Cx(t) + D u(t) +  W_y g \begin{pmatrix} V^T \begin{bmatrix} x(t) \\ u(t) \end{bmatrix} \end{pmatrix},
\end{eqnarray} where $W_x$ and $W_y$ are the linear transformation matrices for transforming the nonlinear polynomial functions in the PNLSS equation and $V$ is the transformation matrix for transforming states and inputs. For completeness of this section, here we describe briefly the algorithm of \cite{Philippe2015Decouple} using the notation of the PNLSS model.  

\begin{figure}[!ht]
\begin{center}
\includegraphics[scale = .325, trim = .5cm 1.5cm 0 1.5cm]{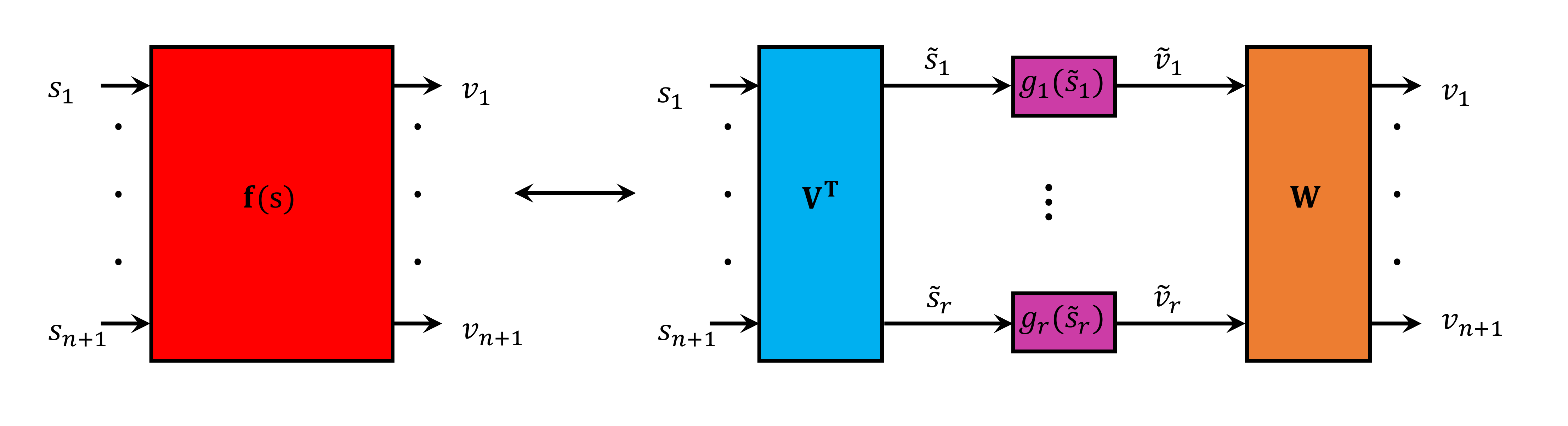}    
\caption{The outcome of the decoupling procedure. A multivariate polynomial vector function is decomposed into a linear transformation $V$, followed by a set of parallel univariate polynomials $g_1,\ldots,g_r$, and another linear transformation $W$. Illustration courtesy \cite{Philippe2015Decouple} } 
\label{fig:CPD}
\end{center}
\end{figure}

Notice that the procedure of Section~\ref{FullNLSS} results in an $n+1$ to $n$ multivariate polynomial vector function (represented by $E$) as well as an $n+1$ to $n_y$ multivariate polynomial vector function (represented by $F$). 
From here on, we treat them jointly as a single $n+n_y$ to $n+1$ multivariate polynomial vector function as 
\begin{equation}
f(x(t),u(t))=\begin{bmatrix}
E \zeta (x(t),u(t))\\
F \eta (x(t),u(t))
\end{bmatrix},
\label{eq:NLPNLSS}
\end{equation}
where $p$ is a function that maps $n+1$ variables to $n+n_y$ outputs. 
Furthermore, let the vector $s$ be defined as
\begin{equation}
s = 
\begin{bmatrix}
x(t) \\
u(t)
\end{bmatrix},
\end{equation}
where $s \in \mathbb{R}^{(n+n_u) \times 1}$. 
We denote by $f(s)$ the original multivariate polynomial function, which is decoupled into $r$ univariate polynomials as 
\begin{equation}
f(s) = W g(V^T s),
\label{eq:fWgVTs}
\end{equation}
where 
\begin{equation}
g\begin{pmatrix} V^T s \end{pmatrix}=
\begin{bmatrix}
g_1(\tilde{s}_1) \\ g_2(\tilde{s}_2) \\ \vdots \\ g_r(\tilde{s}_r),
\end{bmatrix},
\end{equation}
with the univariate functions $g_1(\tilde{s}_1), \ldots, g_r(\tilde{s}_r)$ operating on the transformed variables $\tilde{s}_i = v_i^T s$ and the functions $g_i$ have a fixed user-chosen degree $d$. Fig.~\ref{fig:CPD} shows the result of the decoupling algorithm. Corresponding to the definition of $s$, the matrix $W$ is defined as 
\begin{equation}
W = \begin{bmatrix} W_x \\ W_y \end{bmatrix},
\end{equation}
such that  
\begin{equation}
\begin{bmatrix}
E \zeta (x(t), u(t)) \\
F \eta (x(t), u(t))
\end{bmatrix}
=
\begin{bmatrix}
W_x \\
W_y
\end{bmatrix}
g \begin{pmatrix} V^T \begin{bmatrix}
x(t) \\
u(t)
\end{bmatrix}
 \end{pmatrix}.
\end{equation}
The decoupling method of \cite{Philippe2015Decouple} relies on the fact that the Jacobian of $p$ is given by the expression: 
\begin{equation}
J(s) = W \operatorname{diag}(g_{1}'(v_i^T s), \ldots, g_{r}'(v_r^T s)) V^T,
\label{eq:CPDjac}
\end{equation}
where $v_i$ denotes the $i$th column of $V$.
By considering (\ref{eq:CPDjac}) in a set of sampling points $s^{(k)}$, the question of finding the transformation matrices $V$ and $W$ amounts to solving a simultaneous matrix diagonalization problem, which is essentially what the CPD computes.

\begin{figure}[!ht]
\begin{center}
\includegraphics[trim = 0 1cm 0cm 0, scale = .675]{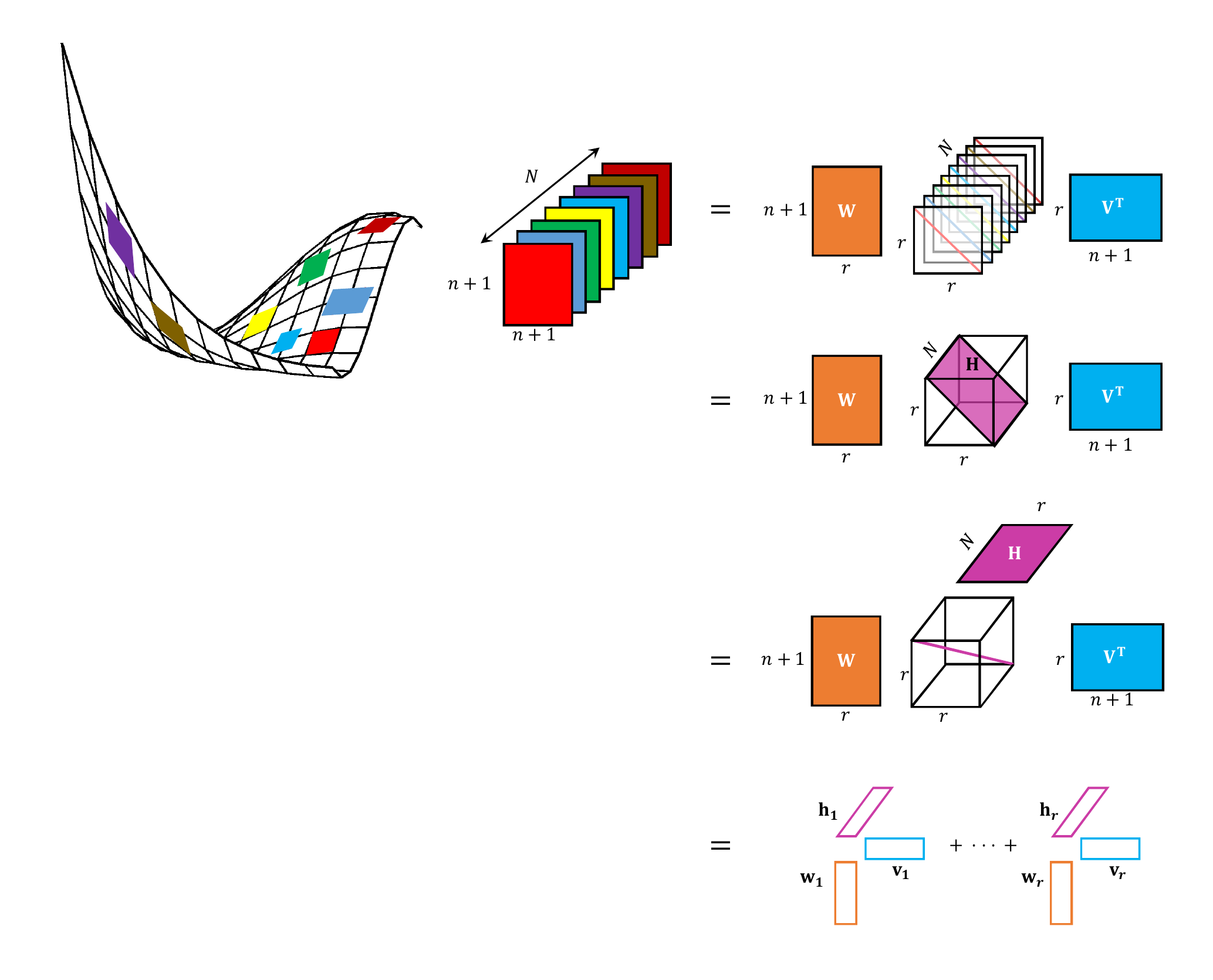}    
\caption{Decoupling multivariate polynomials is done by collecting the Jacobian matrices in a set of sampling points and stacking them into a 3D tensor. The simultaneous matrix diagonalisation of the Jacobian matrices is equivalent to the CPD of this 3D tensor, and returns the transformation matrices V, W and information about the internal functions $g_\ell$. Illustration courtesy \cite{Philippe2015Decouple} }
\label{fig:StackingGrads1}
\end{center}
\end{figure}
\vspace{0cm}

\subsection{Summary of decoupling algorithm}
The decoupling algorithm is adopted from \cite{Philippe2015Decouple} and is briefly summarized here.
\begin{itemize}
\item[1.] The Jacobian of the polynomial vector function $p$ is evaluated in a set of $N_s$ sampling points $s^{(k)}$, for $k = 1,2, \ldots N_s$, which are drawn from a random normal distribution. 
\item[2.] The Jacobian matrices are stacked into an $(n+n_y) \times (n+n_u) \times N_s$ tensor (see Fig.~\ref{fig:StackingGrads1}). 
We have thus 
\begin{equation}
J_{ijk}=\frac{\partial p_i (s_j^{(k)})}{\partial s_j}.
\end{equation}
\item[3.] This tensor is decoupled using the CPD as follows (also see Fig.~\ref{fig:StackingGrads1}) \\
\begin{figure}[!ht]
\begin{center}
\includegraphics[trim = -.5cm 4cm 1cm 0cm, scale = .35 ]{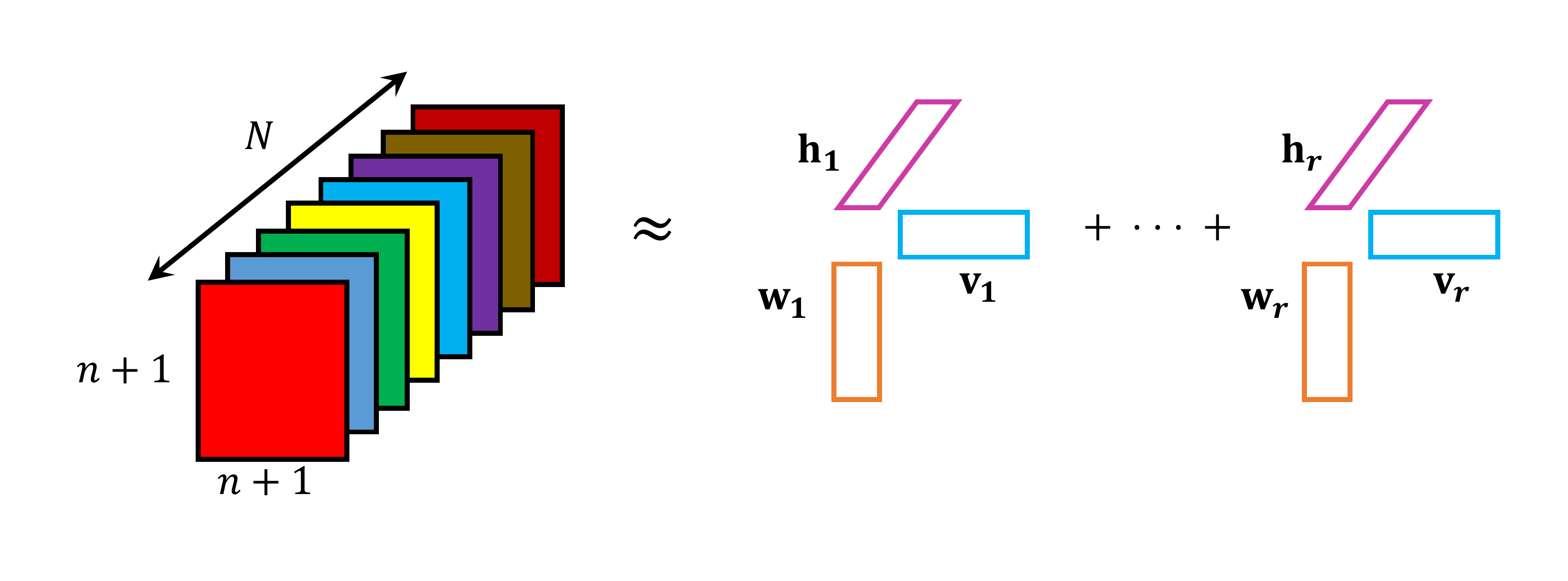}    
\label{fig:StackingGrads2}
\end{center}
\end{figure}
\begin{equation}
J_{ijk} \approx \sum_{\ell = 1}^{r} \, w_{i\ell} \, v_{j\ell} \, h_{k\ell},
\end{equation}
where $W$ and $V$ follow immediately from the CPD and
\begin{equation}
h_{k\ell}= {g_\ell}'(\tilde{s}_\ell^{(k)}),
\end{equation}
is the derivative of the univariate function $g_\ell$ evaluated in sampling points $\tilde{s}_\ell^{(k)}$, for $k = 1, \ldots, N_s$. 
\item[4.] For the $\ell^{th}$ branch ${g}'_\ell(\tilde{s}_\ell)$, we solve the following polynomial fitting
\begin{equation}
\begin{bmatrix}
(\tilde{s}_{i}^{(1)})^1 & (\tilde{s}_{i}^{(1)})^2 & \cdots & (\tilde{s}_{i}^{(1)})^{d-1} \\
(\tilde{s}_{i}^{(2)})^1 & (\tilde{s}_{i}^{(2)})^2 & \cdots & (\tilde{s}_{i}^{(2)})^{d-1} \\
\vdots && \vdots\\
(\tilde{s}_{i}^{(N)})^1 & (\tilde{s}_{i}^{(N_s)})^2 & \cdots & (\tilde{s}_{i}^{(N_s)})^{d-1}
\end{bmatrix}
\begin{bmatrix}
{c}'_{i,2} \\
\vdots \\
{c}'_{i,d-1}
\end{bmatrix}= 
\begin{bmatrix}
{h}_{1i} \\
{h}_{2i} \\
\vdots \\
{h}_{N_{s_{i}}} \\
\end{bmatrix},
\end{equation}
leading to the coefficients of $g'_\ell$.

\item[5.] Solving the symbolic integration
\begin{equation}
g_\ell(\tilde{s}_\ell) = \int {g}'_\ell(\tilde{s}_\ell)d\tilde{s}_\ell,
\end{equation}
determines the functions $g_\ell$ up to the correct value of the integration constants. Notice that, the constant and linear terms are not considered in $g_{\ell}(\tilde{s}_\ell)$. 
\end{itemize}

The result of this algorithm is a decoupled nonlinear state space model with $(2n+n_u+ n_y + d) \times r$ nonlinear parameters. Notice that the number of parameters now increases linearly with the degree as opposed to the combinatorial increase for the multivariate polynomials in \eqref{eq:Comb}. This decoupled model can be used as an initialization for further tuning $V$, $W$ and the coefficients of the $g_\ell$ in order to minimize the output error. Furthermore, the degree $d$ of the nonlinearity can now be increased to capture the nonlinear effects without having to estimate a large number of parameters.

\section{Results}
\label{Results}
The results of the identification procedures described in Section \ref{IdenPNLSS} are presented herein using the benchmark data. The performance of the different model structures is judged based on the following figure of merit.
\begin{equation}
 e_{RMS_t}=\sqrt{\frac{1}{N_t}\sum^{N_t}_{t=1}(y_{mod}(t)-y(t))^2}
\end{equation} where $y_{mod}$ is the modeled output, $y_t$ is the output provided in the test dataset, $N_t$ is the total number of points in $y_t$. 

\subsection{Comparison}
Table \ref{tb:CompModel} shows the comparison between various parameters and hyper-parameters which are associated with different nonlinear model structures explained above.
\begin{table}[!ht]
	\centering
	\caption{Parameters of different nonlinear model structures}
	\vspace{0.2em}
	\begin{tabular}{c @{\hspace{1.0em}} c@{\hspace{0.2em}} c@{\hspace{0.1em}}}
		\hline
		
	\textbf{Model} &	\textbf{Final optimisation parameters} &  \textbf{Hyper-parameters} \\
		\hline
		\hline
		\textbf{BLA} & {A,B,C,D} & {None}\\
		\textbf{PNLSS} & {A,B,C,D,E,F} & {None}\\
		\textbf{PNLSS-I} & {A,B,C,D,E,F,$x_0$} & {None}\\
		\textbf{PNLSS-I$_{DEC}$} & {A,B,C,D,$x_0$,V,W, r, coeff. of $g_1\cdots g_r$} & {$N_s, s^{(k)},d$}\\
		\textbf{NLSS2} & {A,B,C,D, $f_{NL}$, $g_{NL}$} & {$\lambda$, $N_{n}$, bias and weights}\\

		\hline
		\hline
	\end{tabular}
	\label{tb:CompModel}
\end{table}

Two different investigations were performed to study the effect of various factors on performance of the proposed models. For the comparison between BLA and PNLSS model, the benchmark estimation dataset was further divided into model estimation (Model Est $=70$\% of $N_{est} = 1024$ data points) and model validation (Model Val $=30$\% of $N_{est} = 1024$ data points) datasets, whereas the benchmark validation dataset was used as the model test (Model Test $ = N_{val} = 1024$ data points) dataset. For comparison between the different nonlinear model structures original benchmark dataset was used. 

For the estimation of PNLSS and PNLSS-I models, a model order $3$ was selected and the degrees of polynomial basis functions in the state and output equations were set upto degree $3$. Similarly for the PLNSS-I$_{DEC}$ model structure model order was kept fixed at $3$ and $500$ numbers of random data points were selected to obtain the first order information. The number of internal univariate functions was fixed at $5$. For the NLSS2 model structure along with keeping the order of linear part fixed at $3$, the neural-networks with $2$ hidden layer neurons and \textit{sigmoid} nonlinear function were selected to reconstruct the nonlinear functions $f_{NL}$ and $g_{NL}$. 

\begin{table}[!ht]
\centering
\caption{Comparison of the BLA and the PNLSS model}
\begin{tabular}{c @{\hspace{0.55em}} c @{\hspace{0.55em}}c @{\hspace{0.55em}} c @{\hspace{0.25em}} c@{}}
\hline
\textbf{} & \textbf{BLA} &  \textbf{PNLSS} \\
\hline
\hline
\textbf{No. of parameters} & \textbf{$N_{\theta_{BLA}} = 16$} &  \textbf{$N_{\theta_{PNLSS}} = 131$ } \\
\hline
\hline
\textit{\textbf{Model Est}}  & 0.54708 & 0.1133 \\
\textit{\textbf{Model Val}}  & 0.75743 & 0.75063 \\
\textit{\textbf{Model Test}} & 0.75331 & 0.69737 \\
\hline
\end{tabular}
\label{tb:Comp}
\end{table}

\begin{table}[!ht]
	\centering
	\caption{Effect of estimation data and polynomial degree in state equation on the PNLSS-I}
	\begin{tabular}{c @{\hspace{2.75em}} c @{\hspace{2.75em}} c @{\hspace{2.75em}}c @{\hspace{2.75em}} c @{\hspace{0.25em}} c@{}}
		\hline
		\textbf{} &  \textbf{$P_{d_s} = 2$} &  \textbf{$P_{d_s} = 2,3$} \\
		\hline
		\hline
		\textbf{Estimation data length $=500$} \textbf{} &  \textbf{}\\
		\hline
		\textit{\textbf{Model Est}}  & 0.044729 & 0.018701 \\
		\textit{\textbf{Model Val}}  & 0.86985 & 0.87438 \\
		\textit{\textbf{Model Test}} &0.66801 & 0.66404 \\
		\hline
		\hline
		\textbf{Estimation data length  $=600$} & \textbf{} &  \textbf{} \\
		\hline
		\textit{\textbf{Model Est}}  & 0.046441 & 0.020913 \\
		\textit{\textbf{Model Val}}  & 0.88059 & 0.83852 \\
		\textit{\textbf{Model Test}} & 0.67691 & 0.75611 \\
		\hline
		\hline
		\textbf{Estimation data length $=700$} & \textbf{} &  \textbf{} \\
		\hline
		\textit{\textbf{Model Est}}  & 0.041328 & 0.021346 \\
		\textit{\textbf{Model Val}}  & 0.75957 & 0.74075 \\
		\textit{\textbf{Model Test}} & 0.47088 & 0.64381 \\
		\hline
	\end{tabular}
	\label{tb:Comp3}
\end{table}

\begin{table}[!ht]
	\centering
	\caption{Comparison of the nonlinear models}
\begin{tabular}{c @{\hspace{0.65em}} c @{\hspace{0.65em}}c @{\hspace{0.65em}} c @{\hspace{0.25em}} c@{}}
		\hline
		

		\textbf{} &  \textbf{PNLSS-I} &  \textbf{PNLSS-I$_{DEC}$} & \textbf{NLSS2}\\
		\hline
		\hline
		\textbf{No. of parameters} &  \textbf{$N_{\theta_{PNLSS-I}} = 132 $} &  \textbf{$N_{\theta_{PNLSS-I_{DEC}}} = 84 $} & \textbf{$N_{\theta_{NLSS2}}= 71$}\\
		\hline
		\hline
		\textit{\textbf{Benchmark Est}}  & 0.032393 & 0.0324 & 0.1165\\
		\textit{\textbf{Benchmark Val}} & 0.44984  & 0.4972  & 0.3433\\
		\hline
	\end{tabular}
	\label{tb:Comp2}
\end{table} 

\begin{figure}[!ht]
\centering
\includegraphics[width=\textwidth]{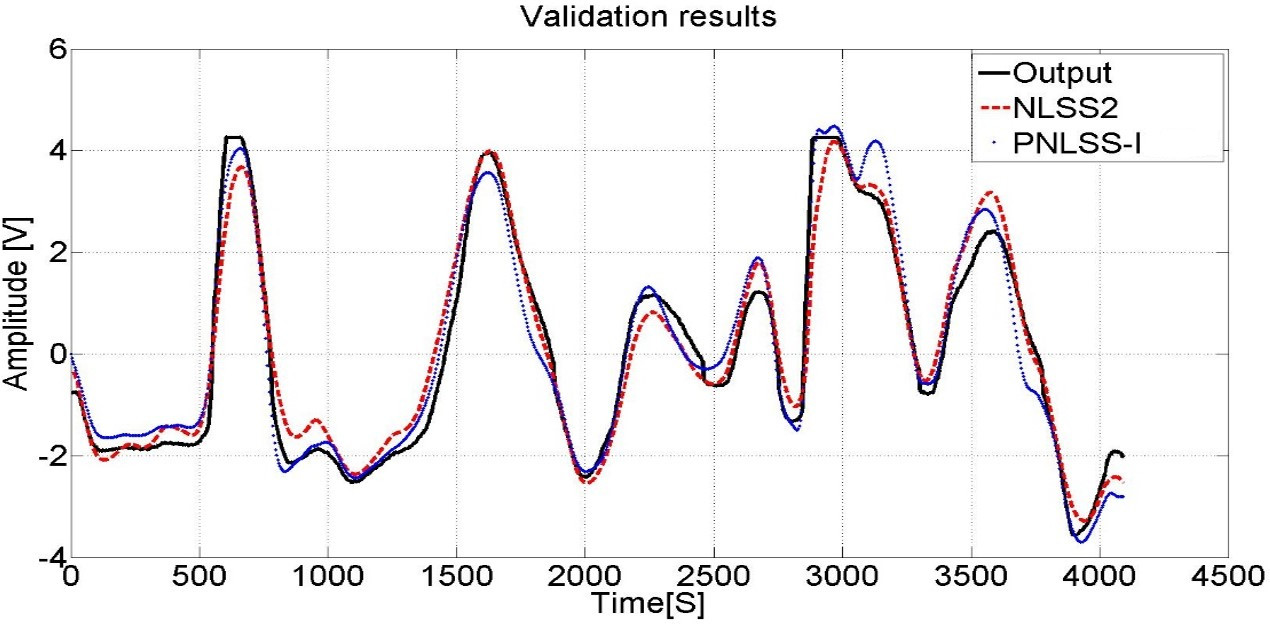}   
\caption{Comparison of the best PNLSS-I and the NLSS2 models on the validation dataset}
\label{ValRes}
\end{figure} 

\begin{figure}[!ht]
\centering
\includegraphics[width=\textwidth]{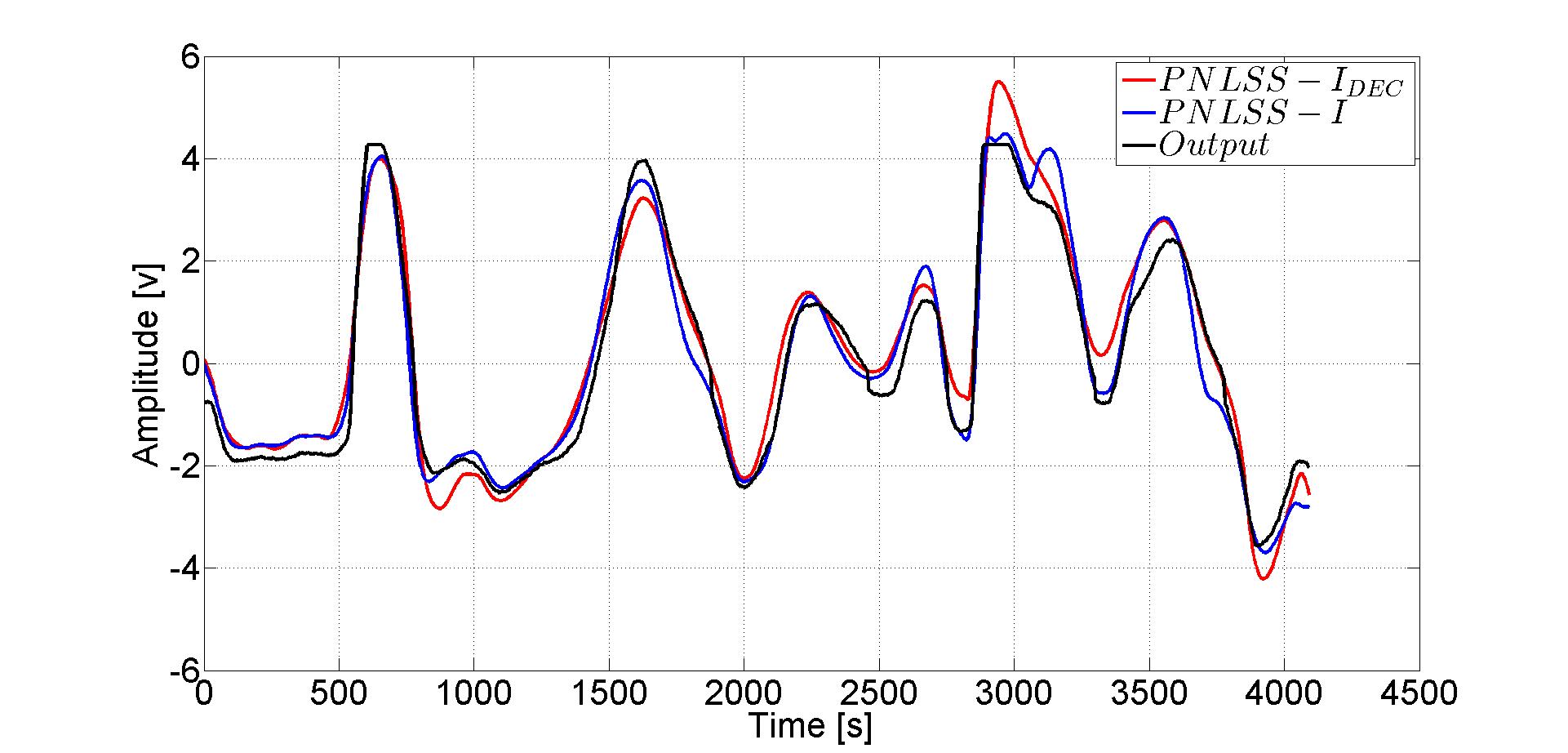}   
\caption{Comparison of  PNLSS-I and PNLSS-I$_{DEC}$ models on the validation dataset}
\label{ValResComp}
\end{figure} 
  
\subsection{Discussion}
Table \ref{tb:Comp} shows the comparison between the BLA and the PNLSS model on the model estimation, model validation and model test datasets respectively. Table \ref{tb:Comp2} shows the comparison between different nonlinear models namely the PNLSS-I, the PNLSS-I$_{DEC}$ and the NLSS2 on the original benchmark dataset. Figure \ref{ValRes} shows the comparison of the outputs of the PNLSS-I and the NLSS2 models with respect to the output provided in the benchmark validation dataset in time domain. 

From Table \ref{tb:Comp}, it is evident that the PNLSS model performs better than the BLA on the model estimation dataset whereas its performance on the model validation and model test datasets is quite similar to the BLA, which point towards the case of overfitting. From Table \ref{tb:Comp2} it can be easily observed that the PNLSS-I model performance is better than the BLA. This is to be expected because PNLSS-I model structure has larger number of data points (full benchmark estimation dataset i.e. $N = 1024$ data points) for the model identification and it estimates the initial state $x_0$ explicitly. But its performance degrades on the benchmark validation dataset too. From Tables \ref{tb:Comp} and \ref{tb:Comp2} it is also clear that, along with larger estimation dataset, once the effect of the initial conditions is taken into account during the identification procedure, the performance of PNLSS-I is also better than the PNLSS model. In the PNLSS model, it is assumed that we start with zero initial conditions and this assumption is violated in the cascaded water-tanks benchmark. 

Table \ref{tb:Comp3} shows the effect of changing the number of estimation data points and the polynomial degree $P_{d_s}$ in the state equation \eqref{eqn:PLNSS} on the performance of the PNLSS-I model. For this test the benchmark estimation dataset was further divided into $500$, $600$, and $700$ data points for the estimation of the PNLSS-I models and the length of validation data points are then $524$, $424$, and $324$ respectively. Although there is no clear visible trend which can be observed from Table \ref{tb:Comp3}, but comparing the results on the estimation and test data, it seems that the model with quadratic and cubic terms overfits, but on the validation data, the model with quadratic and cubic terms performs better than the model with only quadratic terms.

Figure \ref{ValResComp} shows the comparison between PNLSS-I model and PNLSS-I$_{DEC}$ model. It can also be seen from Table \ref{tb:Comp2} that the PNLSS-I$_{DEC}$ model performance on the estimation set remains similar but on the validation dataset it degrades slightly. One of the main reasons for this behavior can be attributed to the fact that the random initialisation of sample points (as well as number of sample points) to obtain the first order information of the multivariate polynomial nonlinear function affects the performance of the decoupled model structure on the validation set. 

Furthermore the number and the degrees of the internal polynomial univariate function also will have an influence on the final output error of the model structure. Even though in the decoupled model, the number of parameters have been reduced to $84$, in this particular benchmark, the number of data points were limited to $1024$, therefore the possibility to choose the degrees of freedom w.r.t above mentioned variables was very limited, hence the performance of PNLSS-I$_{DEC}$ model structure could not be improved further. For a fair comparison and evaluation of different parameters on the performance of PNLSS-I$_{DEC}$, the readers are kindly referred to \cite{Alireza2017}

In the context of this benchmark, NLSS2 model performance is best on the full benchmark validation dataset (see Table \ref{tb:Comp2}). One reason could be due to its better generalisation performance (i.e. less chance of over-fitting because it has less number of parameters $\approx  71$ as compared to PNLSS or PNLSS-I $ \approx 131$ parameters. NLSS2 model structure also offers the possibility to use different basis functions to estimate the nonlinearities in the system as it can be decoupled from the estimation of the dynamics part. Hence the problem of nonlinearity ($f_{NL}(\bullet)$ and $g_{NL}(\bullet)$) estimation can be treated as a static regression problem, which opens up the possibility to use any static function estimation or regression framework to estimate the nonlinear functions $f_{NL}$ and $g_{NL}$.

\section{Conclusion}
\label{Conc}

In this paper, a powerful discrete-time nonlinear state space model structure is introduced to model the nonlinear dynamical systems. The NLSS model is able to capture the dynamic behavior of the system and is capable enough to accommodate any nonlinear effect. It has been shown that this model structure can be initialised very easily. Specifically, in this case study, two different initialisation schemes were introduced, namely PNLSS and NLSS2. 

The NLSS2 performs relatively well on the validation dataset. It has been shown that, it is relatively easy to change the basis functions for estimation of the nonlinearities (nonlinear functions). There are two nonlinearities in this cascaded tank benchmark: a hard saturation due to the overflow and square root nonlinearities in the first principles model. Both nonlinearities can be seen as saturating nonlinearities. The neural networks in the NLSS2 model are better suited than the polynomials in the PNLSS models to model the saturating behaviour. A potential drawback of the NLSS2 method is its computational complexity for large datasets, as was observed in \cite{marconato2014comparison}. We would not recommend to use the NLSS2 method in general, but it is a good option for datasets that are not too large. 
	
PNLSS, PNLSS-I models are a good choice generally for capturing the dynamics of nonlinear system with relatively low state dimensions due to combinatorial increase in the number of parameters. Explicit estimation of initial conditions can easily be done in the PNLSS-I model structure.

A method to obtain the parsimonious representation PNLSS-I$_{DEC}$ of PNLSS model is proposed to deal with the restriction of the PNLSS model structure, in terms of rapid increase in the number of parameters with the increase in the dimension of state-space and inputs. PNLSS-I$_{DEC}$ model can be seen as the model with a dimensionality reduction step to reduce the dimensionality of the multivariate nonlinear polynomial function in PNLSS and PNLSS-I models. This model structure can be a good choice in certain cases where dynamic nonlinearities are involved \cite{Alireza2017}. Finally, promising results were obtained despite having structural limitations in both model structures to deal with the influence of process noise present in the cascaded water-tanks benchmark problem.

\section*{References}
\bibliographystyle{elsarticle-num} 
\bibliography{MSSPBib}
\end{document}